\documentclass[pra,twocolumn,showpacs,groupedaddress,notitlepage,superscriptaddress,floatfix,nofootinbib,preprintnumbers]{revtex4-2}

\usepackage{subfigure}
\usepackage{braket}
\usepackage{tikz}
\usepackage{amsmath}
\usepackage{listings}
\usepackage[ruled]{algorithm} 
\usepackage{algpseudocode}
\usepackage{amsfonts}
\usepackage{float,mathdots}

\usepackage{multirow}
\usepackage{array}
\usepackage{amssymb}
\usepackage{accents}
\usepackage{amsthm}

\usepackage{hyperref}
\usepackage{blkarray}
\usepackage{svg}
\usepackage[dvipsnames]{xcolor}
\usepackage{booktabs}
\usepackage{makecell} 

\usepackage{amsmath}

\usepackage{appendix}

\hypersetup{
    colorlinks=true,   
    linkcolor=cyan,    
    citecolor=magenta, 
    filecolor=magenta, 
    urlcolor=cyan,     
    runcolor=cyan
}
\usepackage[capitalise]{cleveref} 
\newcommand{\appropto}{\mathrel{\vcenter{
  \offinterlineskip\halign{\hfil$##$\cr
    \propto\cr\noalign{\kern2pt}\sim\cr\noalign{\kern-2pt}}}}}

\begin{document}

\preprint{LA-UR-26-28425}

\title{Thermodynamics of Ahn--Doherty--Landahl Continuous Quantum Error Correction}

\author{Juan Garcia-Nila}
\affiliation{Department of Electrical \& Computer Engineering, University of Southern California, Los Angeles, California}

\affiliation{Centre for Quantum Technologies, National University of Singapore, 3 Science Drive 2, 117543, Singapore}
\affiliation{Theoretical Division, Los Alamos National Laboratory, Los Alamos, NM 87545, USA}

\author{Pedro B. Melo}
\affiliation{Departamento de F\'{i}sica, PUC-Rio, 22452-970, Rio de Janeiro RJ, Brazil}
\affiliation{Theoretical Division, Los Alamos National Laboratory, Los Alamos, NM 87545, USA}

\author{Lucas Johns}
\affiliation{Theoretical Division, Los Alamos National Laboratory, Los Alamos, NM 87545, USA}

\begin{abstract}
Continuous quantum error correction (CQEC) replaces discrete syndrome measurements and recovery operations with continuous syndrome extraction and real-time Hamiltonian feedback. Here we investigate the thermodynamic resources required by this process by formulating measurement-based continuous quantum error correction as an information engine. We distinguish the system-side power associated with energy transfer between the feedback field and the protected system from the controller-side power associated with the feedback Hamiltonian. The framework is first developed for the one-qubit Ahn-Doherty-Landahl (ADL) protocol and subsequently extended to the three-qubit repetition code under continuous stabilizer monitoring. Numerical simulations show that increasing the feedback strength improves the steady-state fidelity and reduces the conditional-state entropy, while both energetic contributions continue to increase in magnitude after the fidelity begins to saturate. These results expose a direct tradeoff between logical stabilization
and the energetic resources required for continuous feedback, closely
related to previously established energy--precision tradeoffs in
quantum measurement and quantum-Zeno stabilization.

\end{abstract}

\maketitle

\section{Introduction}

Quantum error correction (QEC) protects quantum information by encoding it in a larger Hilbert space, detecting errors, and applying conditional recovery operations. Conventional QEC organizes these processes into discrete cycles of syndrome extraction and correction. Continuous quantum error correction (CQEC) \cite{Zurek1969,Jacobs2006,Milburn1993,Livingston2022} instead replaces projective measurements and instantaneous gates with continuous weak measurements and finite-strength control fields, establishing a direct connection between error correction and quantum feedback control \cite{Doherty1998,Ahn2002,Chase2008}. Continuous protection can also be implemented autonomously through engineered dissipation or cooling \cite{Sarovar2005,Oreshkov2007,Nila2026}. Recent work has investigated energy–error tradeoffs in discrete, gate-based QEC \cite{Stevens2026}, whereas here we study heat, work, and entropy in continuous measurement-and-feedback error correction.

In the measurement-based protocol introduced by Ahn, Doherty, and Landahl (ADL), continuously acquired syndrome information is used to estimate the conditional quantum state and select a bounded Hamiltonian that increases the code-space fidelity \cite{Ahn2002}. Later work developed more efficient feedback controllers and examined continuous-time error correction under different noise and control assumptions \cite{Chase2008,Lanka2026}. These studies have focused primarily on correction performance, while the energetic requirements of continuously applying and modulating the feedback Hamiltonian remain comparatively unexplored.

The problem lies at the intersection of CQEC and quantum thermodynamics under measurement and feedback. Measurements can change both the state and energy of a quantum system, while feedback converts the acquired information into controlled dynamical action. The energetic and entropic consequences of quantum
measurement have been analyzed within stochastic thermodynamics and
information thermodynamics, including the identification of
measurement-induced energy changes and bounds on the irreversible cost
of generalized measurements~\cite{Elouard2017,
Mancino2018}.

A consistent description must therefore distinguish environmental energy exchange, measurement-induced energy changes, energy transferred to the protected system, and the power associated with modulation of the feedback Hamiltonian \cite{Elouard2017,Prech2025}. Information storage and controller reset also carry thermodynamic consequences, as expressed by Landauer's principle and generalized second laws for feedback-controlled systems \cite{Landauer1961,Sagawa2009}. Related thermodynamic constraints on discrete QEC protocols have connected information gain, heat exchange, and recovery fidelity \cite{Danageozian2022}; however, CQEC requires an explicitly stochastic, continuous-time description.

More recently, the energetic cost of syndrome-memory reset has been
related to the structure of the code, the noise model, and the
representation of the retained syndrome information
\cite{CzartowskiBinder2026}.

Here we formulate measurement-based CQEC as an information engine organized around noise, measurement, feedback, and controller reset. We distinguish the system-side power associated with the feedback-induced change of the bare-system energy from the controller-side power associated with modulation of the feedback Hamiltonian. We first develop the framework for the one-qubit ADL protocol and then extend it to the three-qubit repetition code under continuous stabilizer monitoring.


Our results show that increasing the feedback strength improves stabilization but eventually produces diminishing fidelity gains, while the magnitudes of the feedback-associated energy flows continue to increase. This establishes feedback power as a relevant thermodynamic quantity for assessing and comparing continuous quantum error-correction protocols.

This establishes feedback power as a physical resource for assessing and comparing continuous quantum error-correction protocols.

This paper is organized as follows. In \cref{sec:sec2}, we explain the four thermodynamic strokes for CQEC similar to reference ~\cite{LandiQEC}. We start our exploration with the one-qubit model in \cref{sec:sec3} where we introduce the stochastic master equation, the thermodynamic quantities, and the numerical results. In \cref{sec:sec4} we show the three-qubit repetition code, where we define the stabilizers and the open system dynamics treatment with its thermodynamic observables and the feedback protocol according to ADL \cite{Ahn2002}. In \cref{sec:conclusion} we conclude our main results.

\section{Thermodynamic strokes for Continuous Quantum Error Correction}\label{sec:sec2}

\textit{Quantum error correction} is conventionally described as a sequence of
syndrome measurements followed by conditional recovery operations.
Landi \textit{et al.} formulated this discrete process as an
information engine composed of noise, measurement, feedback, and
memory-reset stages~\cite{LandiQEC}. We briefly review this
thermodynamic framework and adapt it to continuous quantum error
correction (CQEC), in which discrete syndrome measurements and recovery
operations are replaced by continuous weak measurements and real-time
Hamiltonian feedback.

In the discrete setting, the four processes can be represented as
successive strokes of a thermodynamic cycle. In CQEC, however, the
noise, measurement, and feedback processes operate simultaneously and
continuously. They should therefore be understood as conceptually
distinct contributions to the dynamics rather than as temporally
separated strokes. The reset of the information-bearing controller
degrees of freedom completes the thermodynamic description and permits
the continuous-feedback process to operate repeatedly.


\textbf{1. Noise stroke.}
The physical qubit interacts with its environment through decoherence and relaxation processes. In the absence of control, these interactions increase the entropy of the logical state and reduce the logical fidelity.

\textbf{2. Measurement stroke.}
Continuous weak measurements extract information about the system through a stochastic measurement record. This information is stored in the controller and is subsequently used to determine the feedback action.

\textbf{3. Feedback stroke.}
Conditioned on the measurement record, the controller applies a time-dependent Hamiltonian that counteracts the environmental noise. This process transfers energy between the feedback field and the
system while driving the state toward the protected subspace and increases the logical fidelity.

\textbf{4. Reset stroke.}
After the measurement information has been processed, the controller memory must be erased so that the controller memory can be reused in a continued operation fashion. According to Landauer's principle, this reset operation requires a minimum energetic cost proportional to the amount of information stored in the controller.

Unlike conventional quantum heat engines, which are designed to produce work, the objective of the continuous protocol is the protection of quantum information. Here, we characterize the thermodynamic processes that accompany this stabilization.

Consequently, the relevant performance variable is not the extracted work, but in error correction we care about the fidelity of the protected logical state.

\begin{section}{One-Qubit Continuous Error-Correction Model}\label{sec:sec3}

\begin{subsection}{Stochastic Master Equation}

We first consider the simplest setting, the one-qubit continuous feedback protocol introduced by Ahn, Doherty, and Landahl (ADL). Although the model describes only a single physical qubit, it captures all essential ingredients of continuous quantum error correction: continuous syndrome acquisition, stochastic conditional dynamical evolution and Hamiltonian feedback.

The physical system qubit is described by the Hamiltonian
\begin{equation}\label{eq:HS_1q}
    H_S=\frac{\Delta}{2}(I-Z),
\end{equation}
whose ground state $\ket{0}$ is meant to be protected so it defines the desired logical state. 

Bit-flip errors are modeled through the Lindblad operator $\mathcal{L}_{\rm noise}=\sqrt{\gamma} \sigma_x$, while continuous weak monitoring of the logical observable is performed through a measurement of the Pauli operator $\sigma_z$ with measurement strength $\kappa$.

The conditional evolution of the qubit is governed by the stochastic master equation

\begin{align}\label{eq:SME_1q}
d\rho=&-i[H_S+H_{\rm fb},\rho]dt+\gamma \mathcal{D}[\sigma_x]\rho dt \nonumber \\
&+\kappa \mathcal{D}[\sigma_z]\rho dt +\sqrt{\eta \kappa}\mathcal{H}[\sigma_z]\rho dW,
\end{align}
where the dissipation term is
\begin{equation}
    \mathcal{D}[L]\rho=L\rho L^{\dagger}-\frac{1}{2}\{L^{\dagger}L,\rho\}
\end{equation}
and the measurement back-action term
\begin{equation}
    \mathcal{H}[L]\rho=L\rho+\rho L^{\dagger}-\mathrm{Tr}[(L+L^{\dagger})\rho]\rho.
\end{equation}
The Wiener increment satisfies $dW^2=dt$ and represents the stochastic measurement backaction.

Following \cite{Ahn2002}, the controller applies the feedback Hamiltonian
\begin{equation}\label{eq:Hfb_1q}
    H_{\rm fb}(t)=\frac{\Omega}{2}u(t)\sigma_x
\end{equation}
where $\Omega$ denotes the maximum feedback strength and $u(t)$ is a bounded control signal satisfying $|u(t)|\leq 1$. In the ADL protocol, 
\begin{equation}
    u(t)=\mathrm{sgn}(\langle Y\rangle)
\end{equation}
so that the controller continuously rotates the Bloch vector toward the target state.

The conditional density matrix is conveniently written in Bloch form,
\begin{equation}\label{eq:Bloch_1q}
    \rho(t)=\frac{1}{2}(I+x(t)\sigma_x+y(t)\sigma_y+z(t)\sigma_z),
\end{equation}
from which all thermodynamic quantities considered in this work will be derived.
\end{subsection}

\begin{subsection}{Thermodynamic description}

Now that we have introduced the stochastic dynamics of a continuously monitored qubit, we formulate its thermodynamic description. Throughout this work we distinguish three fundamentally different physical processes:

1. the exchange of energy between the physical qubit and its environment,

2. the work performed by the feedback controller,

3. the storage of information through continuous measurement.

This separation allows us to formulate continuous quantum error correction as a genuine nonequilibrium thermodynamic process.

\begin{subsubsection}{Internal Energy}

The internal energy of the physical qubit, defined with respect to the system Hamiltonian \cref{eq:HS_1q}, is
\begin{equation}\label{eq:Int_Ene}
    U(t)\equiv \mathrm{Tr}[H_S\rho(t)].
\end{equation}
Using the Bloch decomposition \cref{eq:Bloch_1q} it can be rewritten as
\begin{equation}
    U(t)=\frac{\Delta}{2}(1-z(t)).
\end{equation}
The logical fidelity with respect to the target state is
\begin{equation}
    F(t)=\frac{1+z(t)}{2},
\end{equation}
so the internal energy could be written as
\begin{equation}
    U(t)=\Delta (1-F(t)).
\end{equation}
This simple relation shows that logical protection has an immediate energetic interpretation: increasing the fidelity corresponds to reducing the internal energy of the physical qubit.
\end{subsubsection}

\begin{subsubsection}{First Law of Thermodynamics}

The evolution of the internal energy follows directly from the stochastic master equation,
\begin{equation}
    dU=\mathrm{Tr}[H_S d\rho],
\end{equation}
because $\rho$ follows an Itô stochastic equation and generally has no ordinary time derivative.

Since the conditional evolution contains Hamiltonian, dissipative, and measurement contributions, the first law naturally decomposes into
\begin{equation}\label{eq:first_law_t}
dU=\delta Q+\delta W_{\rm sys}
\end{equation}
where $\delta Q$ collects the energy changes generated by the environmental noise and continuous measurement, $\delta W_{\rm sys}$ represents the energetic exchange generated by the noncommutativity of the system and feedback Hamiltonians.

Each contribution has a distinct physical origin and should not be interpreted interchangeably.
\end{subsubsection}

\begin{subsubsection}{System Power}
    Since the feedback Hamiltonian \cref{eq:Hfb_1q} does not commute with the system Hamiltonian \cref{eq:HS_1q}, there is a continuous exchange of energy with the physical qubit.

The system power is defined as the feedback-induced rate of change of
the bare-system energy. Since $U_S(t)=\mathrm{Tr}[H_S\rho(t)]$ and the
feedback contribution to the state evolution is
$d\rho|_{\rm fb}=-i[H_{\rm fb}(t),\rho(t)]dt$, cyclicity of the trace
gives
\begin{equation}\label{eq:P_sys}
    P_{\rm sys}(t)
    \equiv \frac{dU_S}{dt}\bigg|_{\rm fb}
    =-i\mathrm{Tr}\!\left\{[H_S,H_{\rm fb}(t)]\rho(t)\right\}.
\end{equation}
Thus, $P_{\rm sys}$ quantifies the instantaneous energy transferred
to the bare system through the feedback-induced evolution of its state.

For the one-qubit code, the commutator between the system \cref{eq:HS_1q} and feedback \cref{eq:Hfb_1q} Hamiltonians is
\begin{equation}
    [H_S,H_{\rm fb}]=-i\frac{\Delta \Omega}{2}\sigma_y,
\end{equation}
yielding
\begin{equation}
    P_{\rm sys}=-\frac{\Delta \Omega}{2}uy.
\end{equation}

Since the power depends on each trajectory, we can define an asymptotic time average of the system power
\begin{equation}
    P_{\rm sys}^{\infty}=\lim_{T\rightarrow \infty} \frac{1}{T}\int_0^T P_{\rm sys}(t)dt.
\end{equation}

Using the exact fidelity equation derived from the Bloch dynamics, the exact conditional fidelity SDE is
\begin{equation}
dF=\bigg[\frac{\Omega}{2}uy-\gamma(2F-1)\bigg]dt+\sqrt{\eta \kappa}(1-z^2)dW,
\end{equation}

This relation establishes a direct connection between energetic expenditure and logical-state evolution.
\end{subsubsection}

\begin{subsubsection}{Feedback Power}

The previous contribution describes energy transferred between the feedback Hamiltonian and the physical qubit while the Hamiltonian itself remains fixed.

A second energetic contribution arises because the controller continuously changes the Hamiltonian in response to the measurement record.

This controller-side contribution does not enter in \cref{eq:first_law_t}, because the internal energy \(U\) is defined with respect to the bare-system Hamiltonian \(H_S\). It nevertheless quantifies the external work associated with implementing the feedback protocol.

Following the standard definition of externally performed work,
\begin{equation}\label{eq:P_fb}
    P_{\rm fb}(t)=\mathrm{Tr}\left(\rho \dot{H}_{\rm fb}\right).
\end{equation}


Similarly, we define the asymptotic average feedback power as the average feedback power

\begin{equation}
    P_{\rm fb}^{\infty}=\lim_{T\rightarrow \infty} \frac{1}{T}\int_0^T P_{\rm fb}(t)dt.
\end{equation}

Unlike the system work, this quantity represents energy supplied directly by the external controller through modulation of the control Hamiltonian. It therefore characterizes the energetic cost of implementing the feedback protocol itself rather than the energy exchanged with the physical qubit.

With our sign convention, negative power corresponds to energy extraction, and we use the magnitude of the power when discussing the energetic resources required by the protocol.

Although $P_{\rm sys}$ and $P_{\rm fb}$ originate from the same feedback process, they quantify fundamentally different thermodynamic resources.
\end{subsubsection}
\end{subsection}

\subsubsection{Conditional-State Entropy}

In our continuous protocol, information about the system is continuously
extracted through weak measurements and used by the controller to
generate a feedback correction. The uncertainty remaining in the quantum
state, conditioned on the measurement record available to the
controller, is quantified by the von Neumann entropy
\begin{equation}\label{eq:entropy}
    S(\rho_c)=-\mathrm{Tr}(\rho_c\ln\rho_c).
\end{equation}
For the one-qubit conditional state written in the Bloch representation
in \cref{eq:Bloch_1q}, the eigenvalues are
\begin{equation}\label{eq:e_vals_1q}
    \lambda_{\pm}=\frac{1\pm r}{2},
\end{equation}
where
\begin{equation}
    r=\sqrt{x^2+y^2+z^2}.
\end{equation}
The entropy is therefore
\begin{equation}
    S(\rho_c)
    =
    -\lambda_+\ln\lambda_+
    -\lambda_-\ln\lambda_-.
\end{equation}

Unlike the logical fidelity, which depends only on the longitudinal
Bloch component $z$, the entropy depends on the length of the complete
Bloch vector. It therefore characterizes the uncertainty remaining in
the conditional state given the continuous measurement record.

Although the system is initially prepared in a pure state, the
conditional state does not generally remain pure because the
environmental bit-flip channel is not directly observed. Over a short
time interval, this channel produces
\begin{equation}
    \rho_c(t+dt)
    \simeq
    (1-\gamma dt)\rho_c(t)
    +\gamma dt\,X\rho_c(t)X.
\end{equation}
Since the controller does not know whether a bit flip occurred, its
conditional state becomes a mixture of the flipped and unflipped
possibilities. This can also be seen from the purity
\begin{equation}
    \mathcal{P}(t)
    =
    \mathrm{Tr}[\rho_c^2(t)]
    =
    \frac{1+x^2+y^2+z^2}{2},
\end{equation}
whose rate of change under the unobserved bit-flip channel is decreased
\begin{equation}
    \left.\frac{d\mathcal{P}}{dt}\right|_{\rm noise}
    =
    -2\gamma(y^2+z^2)\leq0.
\end{equation}

Continuous measurement plays the opposite informational role. The
measurement record provides information about the possible error and consequently it increases the conditional purity. For
unit measurement efficiency, the monitored channel by itself preserves
the purity of a state conditioned on its complete measurement record.
Nevertheless, the conditional state in the present model can remain
mixed because the independent bit-flip channel is unobserved.

The feedback Hamiltonian generates unitary evolution and therefore
does not directly change the eigenvalues or entropy of the
instantaneous conditional state. Instead, it uses the information
acquired through continuous monitoring to rotate the state toward the
protected subspace. Thus, the steady-state entropy results from the
competition between mixing caused by the noise, information
acquisition through measurement and the corrective Hamiltonian
feedback.

Finally, the conditional von Neumann entropy $S(\rho_c)$ should not be
identified with the classical entropy stored in the controller memory.
The former quantifies the controller's remaining uncertainty about the
quantum state, whereas the latter depends on how the measurement record
is sampled and stored.

\begin{figure*}[t]
\centering

\begin{minipage}[t]{0.46\textwidth}
    \centering
\includegraphics[width=\linewidth]{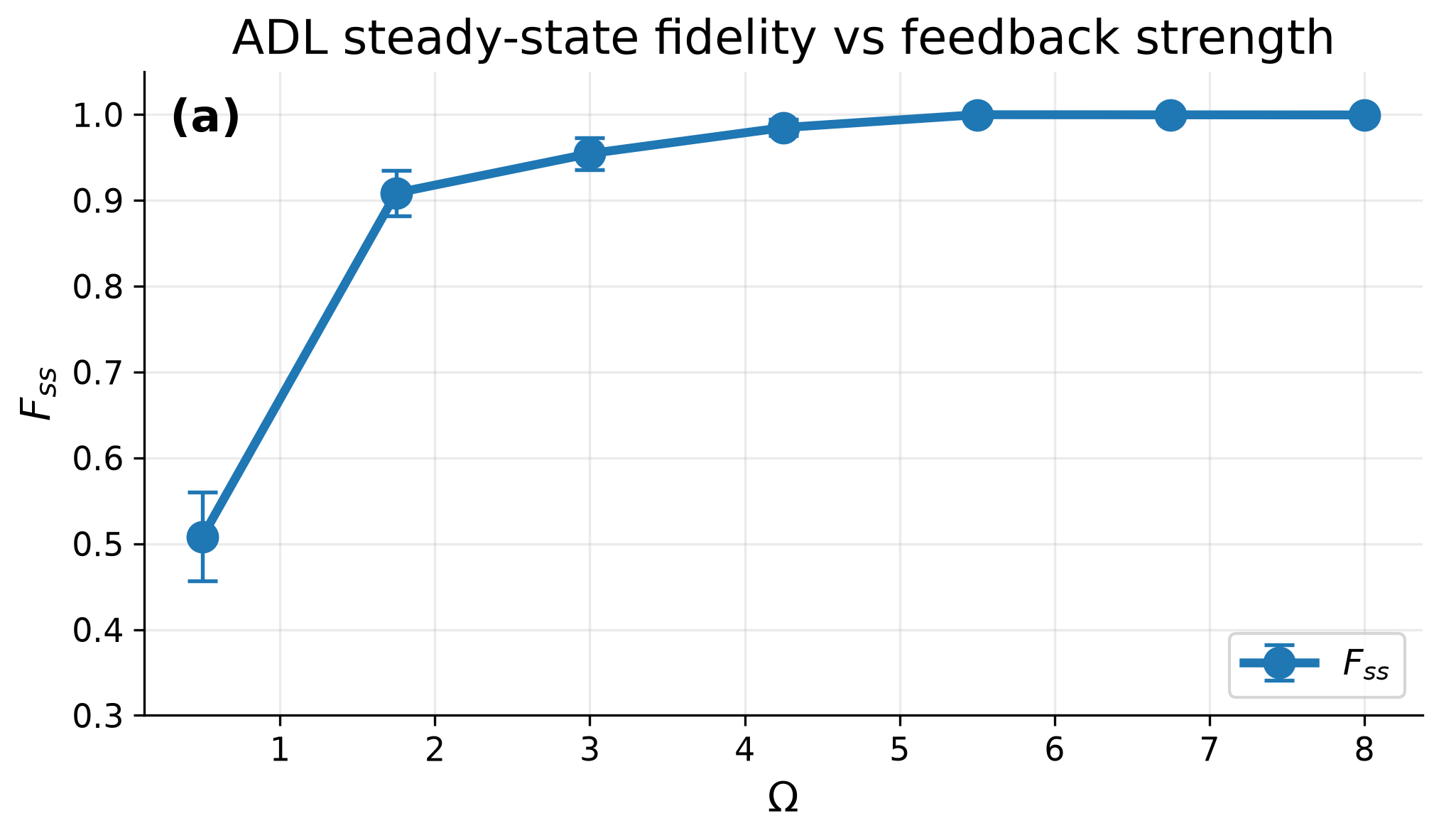}
\end{minipage}\hfill
\begin{minipage}[t]{0.46\textwidth}
    \centering
\includegraphics[width=\linewidth]{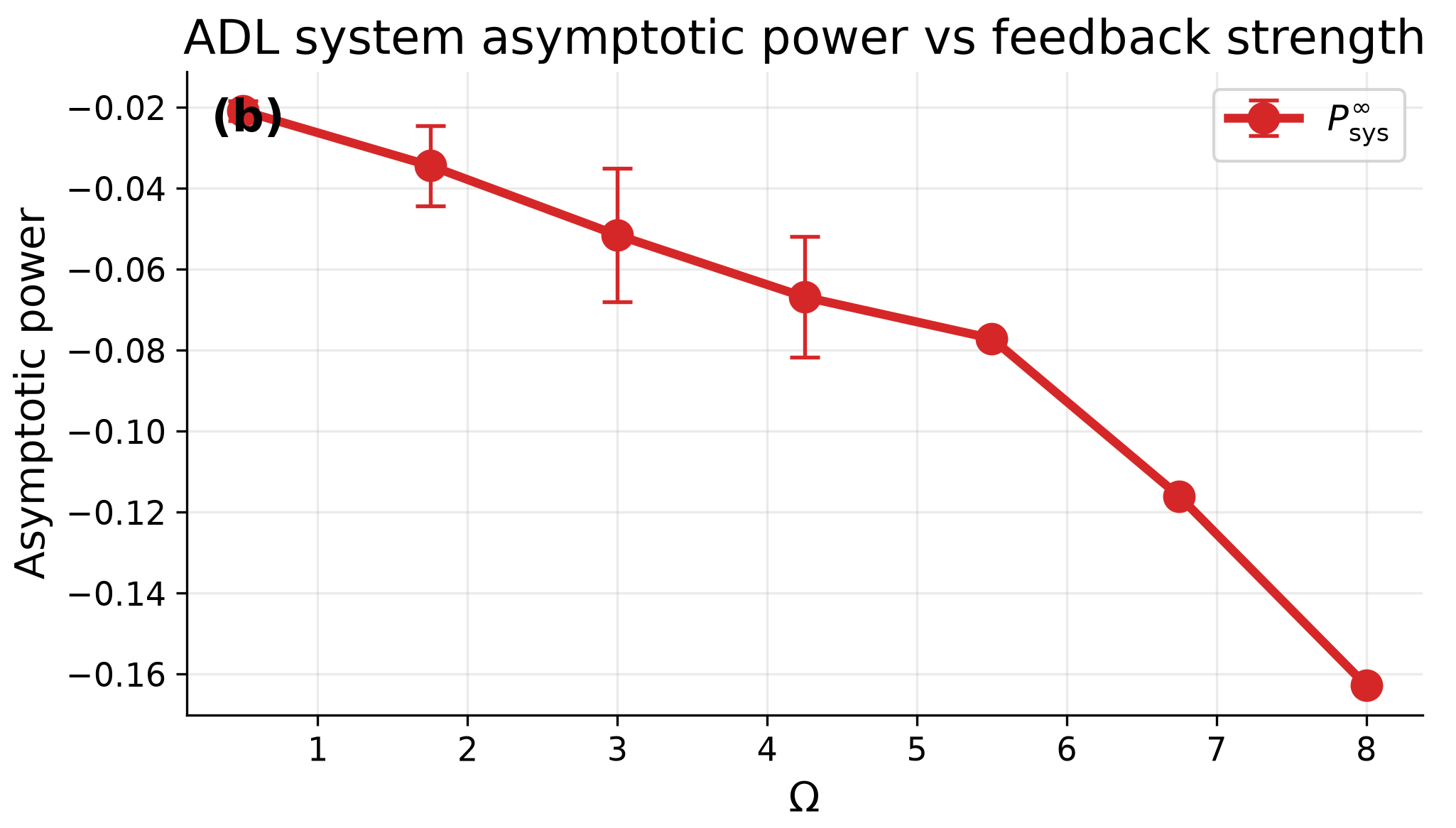}
\end{minipage}
\vspace{0.4cm}
\begin{minipage}[t]{0.46\textwidth}
    \centering
\includegraphics[width=\linewidth]{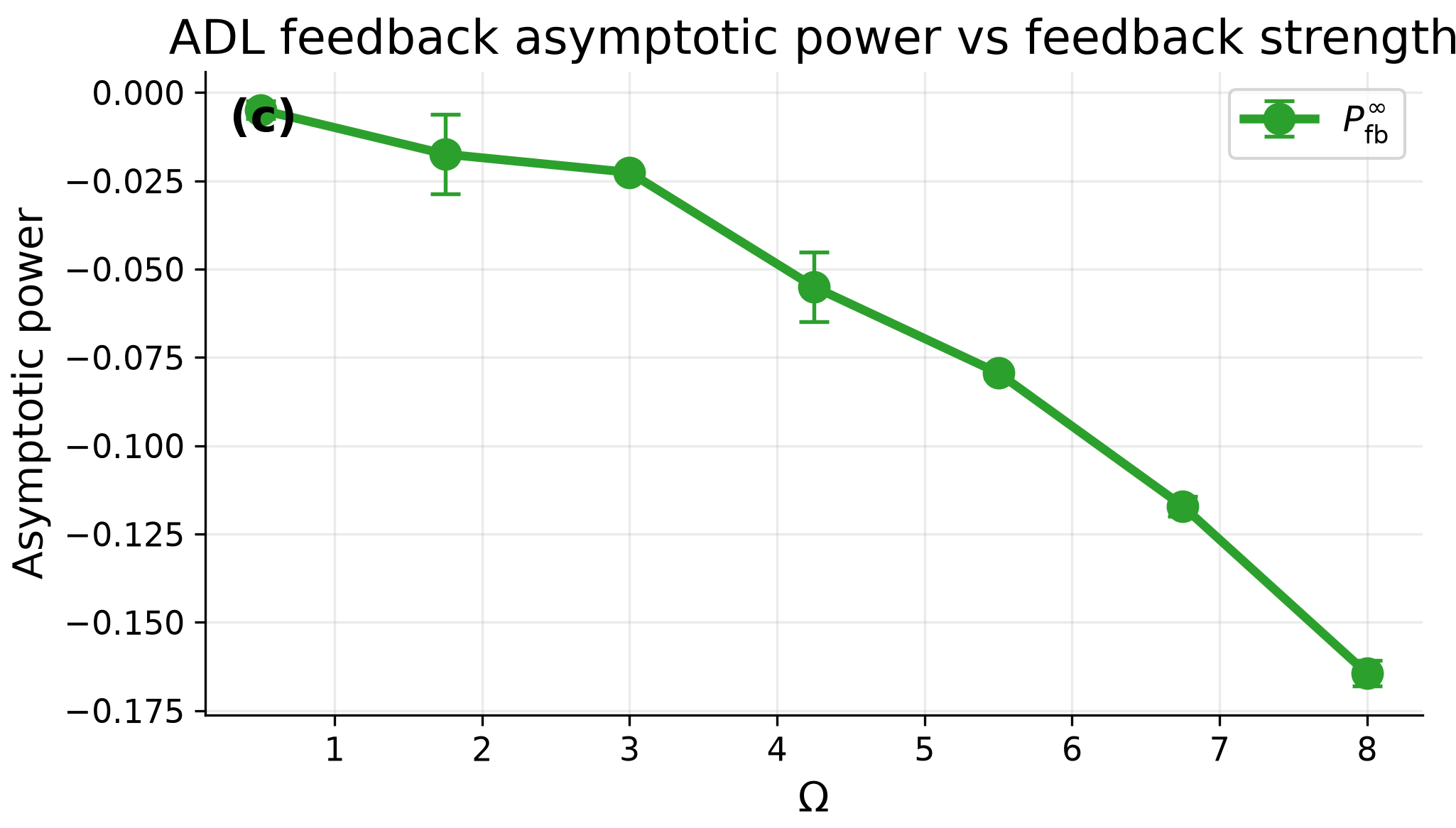}
\end{minipage}\hfill
\begin{minipage}[t]{0.46\textwidth}
    \centering
\includegraphics[width=\linewidth]{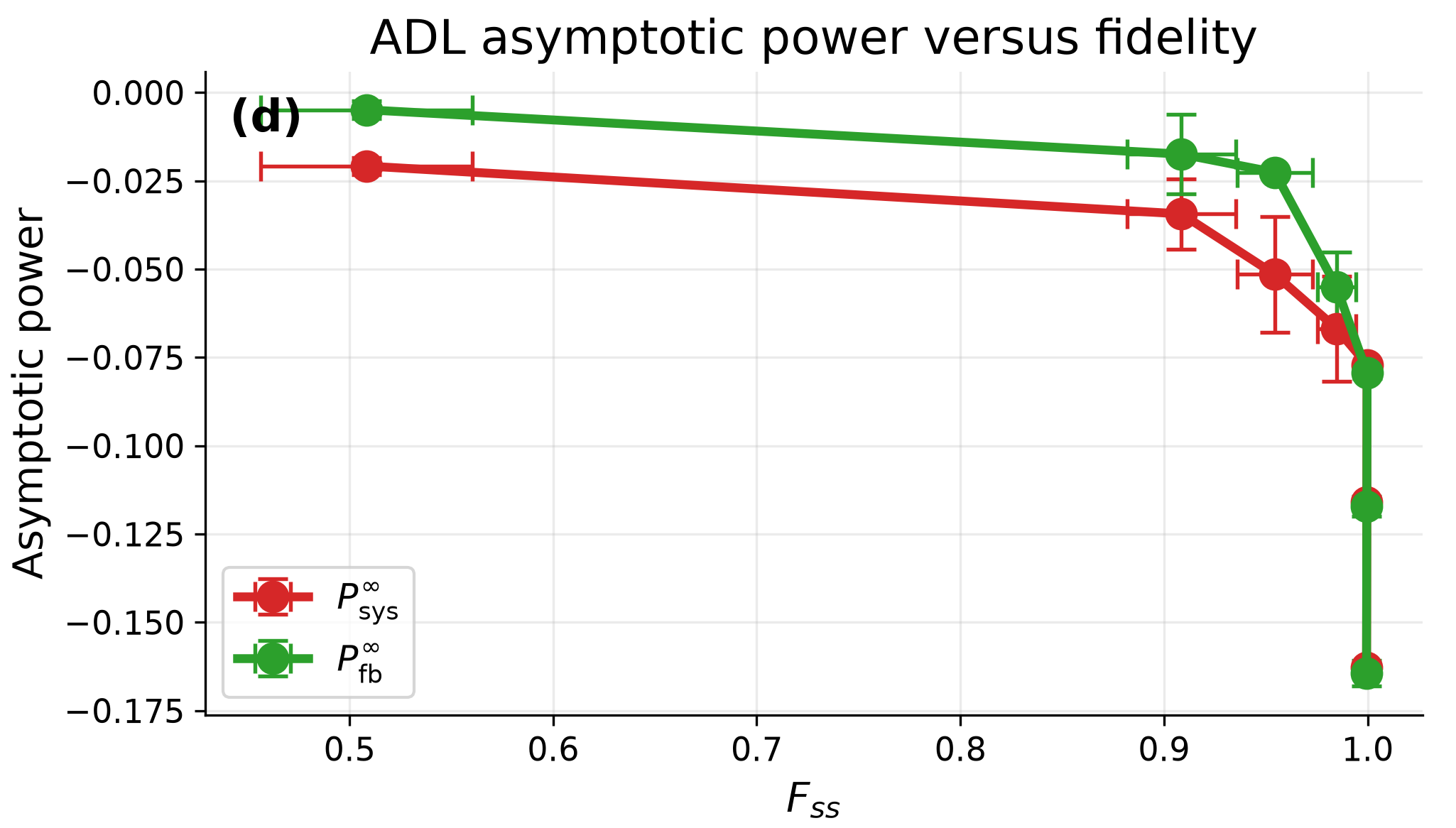}
\end{minipage}
\vspace{0.4cm}
\begin{minipage}[t]{0.46\textwidth}
    \centering
\includegraphics[width=\linewidth]{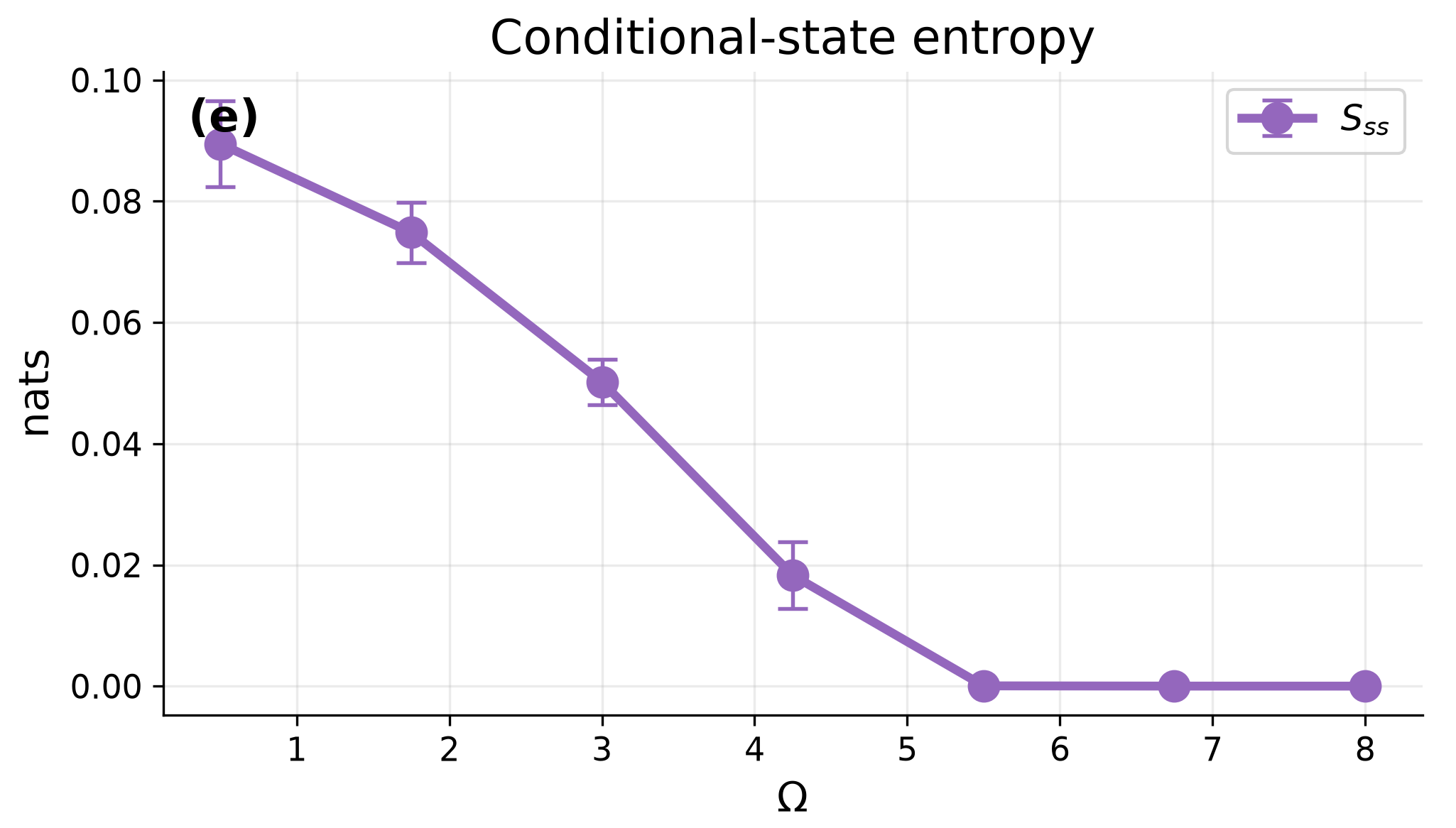}
\end{minipage}

\caption{Thermodynamic properties of the one-qubit ADL continuous feedback protocol as a function of the feedback strength.
(a) Steady-state logical fidelity, (b) asymptotic system power $P_{\rm sys}^{\infty}(T)$, (c) feedback work $P_{\rm fb}^{\infty}(T)$, (d) asymptotic average power versus logical fidelity and (e) von Neumann conditional-state entropy.
}
\end{figure*}

\begin{subsection}{Numerical results for one-qubit code}

To illustrate the thermodynamic framework developed in the previous section, we numerically solve the stochastic master equation \cref{eq:SME_1q} for different feedback strengths $\Omega$. For each value of $\Omega$, we compute the steady-state logical fidelity together with the asymptotic average system power, feedback power, and entropy.

Explicitly for the bit flip noise, the heat exchange is
\begin{equation}
    \delta Q = \Delta \gamma z_t dt,
\end{equation}
the internal energy change
\begin{equation}
dU=\Delta \gamma z dt - \frac{\Delta \Omega}{2}uydt-\Delta \sqrt{\eta \kappa}(1-z^2)dW.
\end{equation}
and the feedback power
\begin{equation}
    P_{\rm fb}=\frac{ \Omega}{2}x(t)\dot{u}(t),
\end{equation}
where, because we use bang-bang feedback, $\dot u$ is evaluated with the finite timestep and hysteresis window used numerically.

A more microscopic description would introduce an explicit detector
and controller with finite response time. Finite-bandwidth continuous
feedback can be described through joint system--detector dynamics,
which naturally regularizes instantaneous switching and incorporates
the delay and noise of the measurement apparatus
\cite{AnnbyAndersson2022}.

\begin{subsubsection}{Fidelity}
Figure 1 summarizes the energetic behavior of the ADL feedback protocol. Figure 1(a) shows that the steady-state logical fidelity increases monotonically with the feedback strength and rapidly approaches unity. Beyond $\Omega \approx 5$, however, the fidelity enters a saturation regime where further increases in the control amplitude produce only marginal improvements. This behavior indicates that the controller is already able to effectively suppress the stochastic bit-flip errors, and additional control provides progressively smaller gains.
\end{subsubsection}
\begin{subsubsection}{Work}
The corresponding power is shown in Figs. 1(b) and 1(c). In contrast with the logical fidelity, both the asymptotic  system power $P_{\rm sys}$ and the feedback power $P_{\rm fb}$ continue to increase in magnitude as the feedback strength grows. Stronger feedback therefore requires a continuously increasing energetic investment even after the logical fidelity has nearly reached its maximum value.

This contrast is emphasized in Fig.~1(d), where the power is plotted
directly as a function of the steady-state fidelity. 

Eliminating the control parameter reveals a direct relation between
the asymptotic powers and the steady-state fidelity: the powers vary
more strongly as the fidelity approaches its saturation regime.
This behavior is qualitatively reminiscent of previously studied
relations between energy and precision in quantum measurement and
quantum-Zeno stabilization~\cite{Abdelkhalek2018}.

\end{subsubsection}
\begin{subsubsection}{Conditional-state entropy}

The information-theoretic interpretation of the protocol is presented in Fig. 1(e). As the feedback strength increases, the von Neumann entropy of the conditional state decreases monotonically towards zero, indicating that the controller continuously purifies the qubit despite the presence of environmental noise. 

Together, these results show that increasing the feedback strength
improves the logical fidelity while reducing the uncertainty remaining
in the conditional state. This entropy reduction results from the
interplay between information acquisition through continuous
measurement and corrective Hamiltonian feedback.

\end{subsubsection}
\end{subsection}
\end{section}

\begin{section}{Three-Qubit Repetition Code}\label{sec:sec4}

The one-qubit model introduced in the previous section is useful as a minimal thermodynamic benchmark, but it does not yet constitute a genuine error-correcting code. We therefore now extend the analysis to the three-qubit repetition code, which is the simplest stabilizer code capable of continuously detecting and correcting single-qubit bit-flip errors. In contrast with the one-qubit protocol, where feedback acts directly on a single physical qubit, the repetition code protects a logical qubit encoded nonlocally across three physical qubits. As a result, the monitored observables, the feedback Hamiltonian, and the definition of fidelity must all be reformulated in the code space.

\begin{subsection}{Code space and stabilizer generators}
We take as logical basis states 
\begin{equation}
   \ket{0_L}=\ket{000},\qquad \ket{1_L}=\ket{111},
\end{equation}
so that the codespace uses the majority vote to detect errors. Two corresponding independent stabilizer generators may be chosen as
\begin{equation}
    S_1=Z_1Z_2,\, S_2=Z_2Z_3 
\end{equation}
In the numerical implementation it is therefore sufficient to monitor the two independent stabilizers for the codespace projector and in the interpretation of the code-space fidelity.

The logical fidelity of the repetition code is defined as
\begin{equation}
    F_{\rm code}(t)=\mathrm{Tr}[P_{\rm code}\rho(t)].
\end{equation}
Unlike the one-qubit fidelity, which is simply the population of the protected ground state, $F_{\rm code}$ measures the overlap of the conditional state with the entire encoded logical subspace.
\end{subsection}

\begin{subsection}{Open system dynamics}
The physical noise model consists of independent bit-flip errors acting on each of the three qubits. The system Hamiltonian is chosen as
\begin{equation}
    H_S=\frac{\Delta}{2}\sum_{i=1}^3 (I-Z_i),
\end{equation}
which is the natural three-qubit generalization of the one-qubit Hamiltonian used earlier. With this choice, the internal energy is simply proportional to the total excitation number. The two logical basis states are not
energetically degenerate: $\ket{000}$ has energy zero, whereas
$\ket{111}$ has energy $3\Delta$. Consequently, this bare-system
energy is not determined solely by the code-space fidelity.

The conditional state evolves according to a stochastic master equation of the form

\begin{align}
    d\rho=&-i[H_S+H_{\rm fb}(t),\rho]dt+\gamma \sum_{i=1}^3 \mathcal{D}[X_i]\rho dt
   \nonumber \\
&+\kappa\sum_{a=1}^2\mathcal{D}[S_a]\rho dt+\sqrt{\eta \kappa}\sum_{a=1}^2 \mathcal{H}[S_a]\rho dW_a.
\end{align}

Here $\kappa$ is the strength of the continuous stabilizer measurements, $\eta$ is the measurement efficiency, and $dW_1$, $dW_2$ are independent Wiener increments associated with the two monitored syndrome channels. This is the main structural difference with the one-qubit model: the monitored observables are now stabilizer parities rather than a single Bloch-component observable like in the one-qubit code.
\end{subsection}

\begin{figure*}[t!]
\centering

\begin{minipage}[t]{0.46\textwidth}
    \centering
\includegraphics[width=\linewidth]{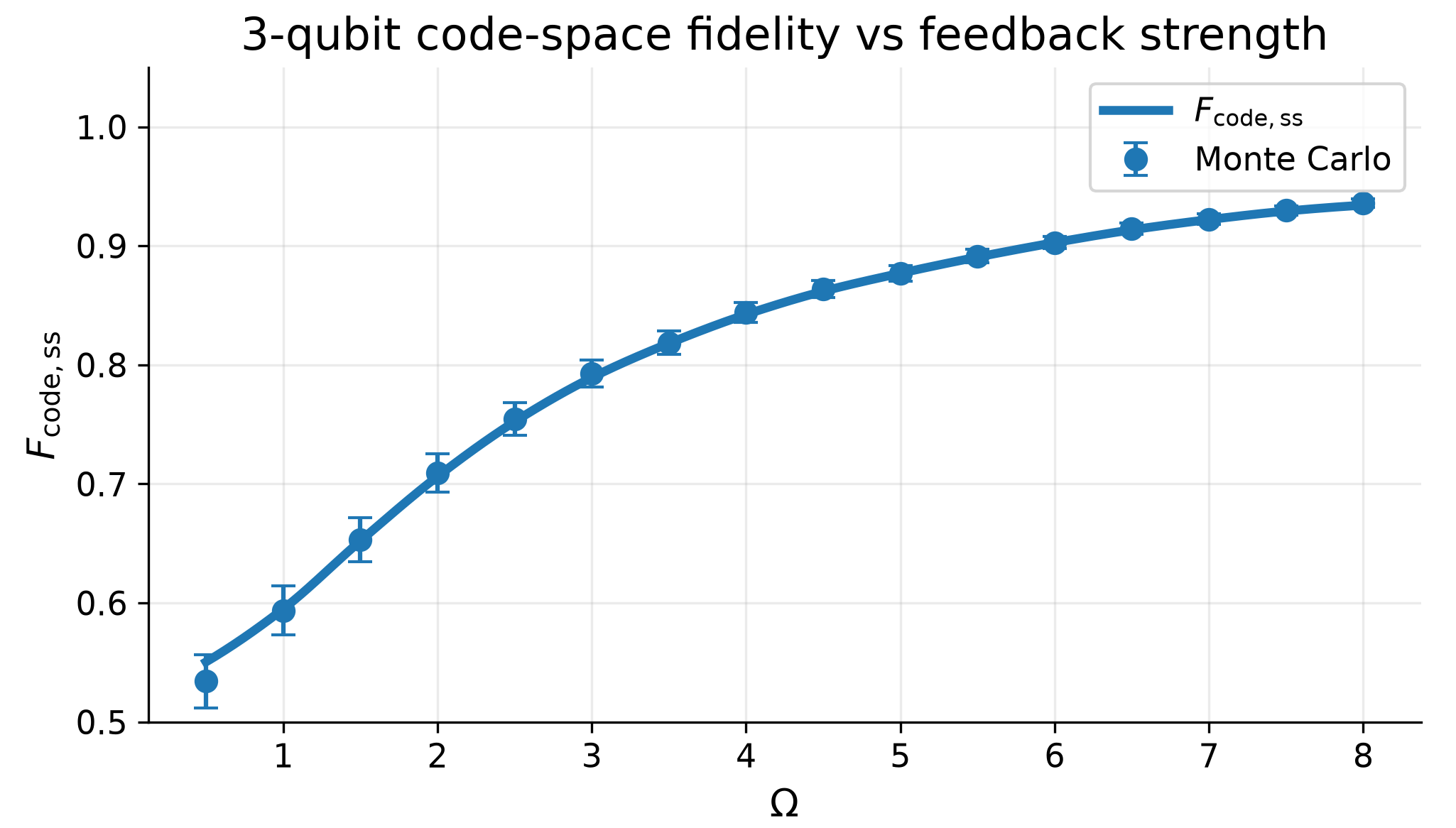}
\end{minipage}\hfill
\begin{minipage}[t]{0.46\textwidth}
    \centering
\includegraphics[width=\linewidth]{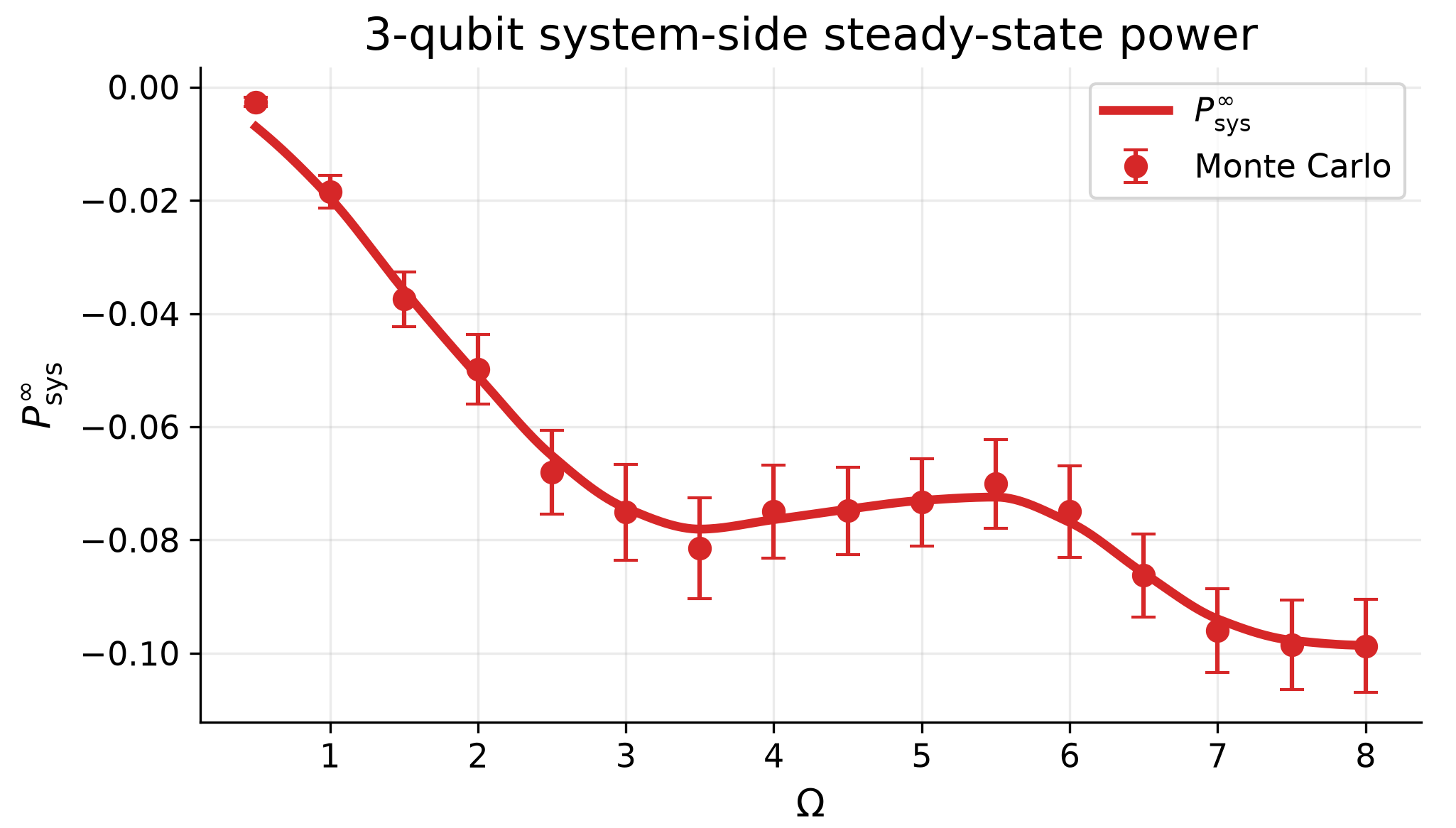}
\end{minipage}
\vspace{0.4cm}
\begin{minipage}[t]{0.46\textwidth}
    \centering
\includegraphics[width=\linewidth]{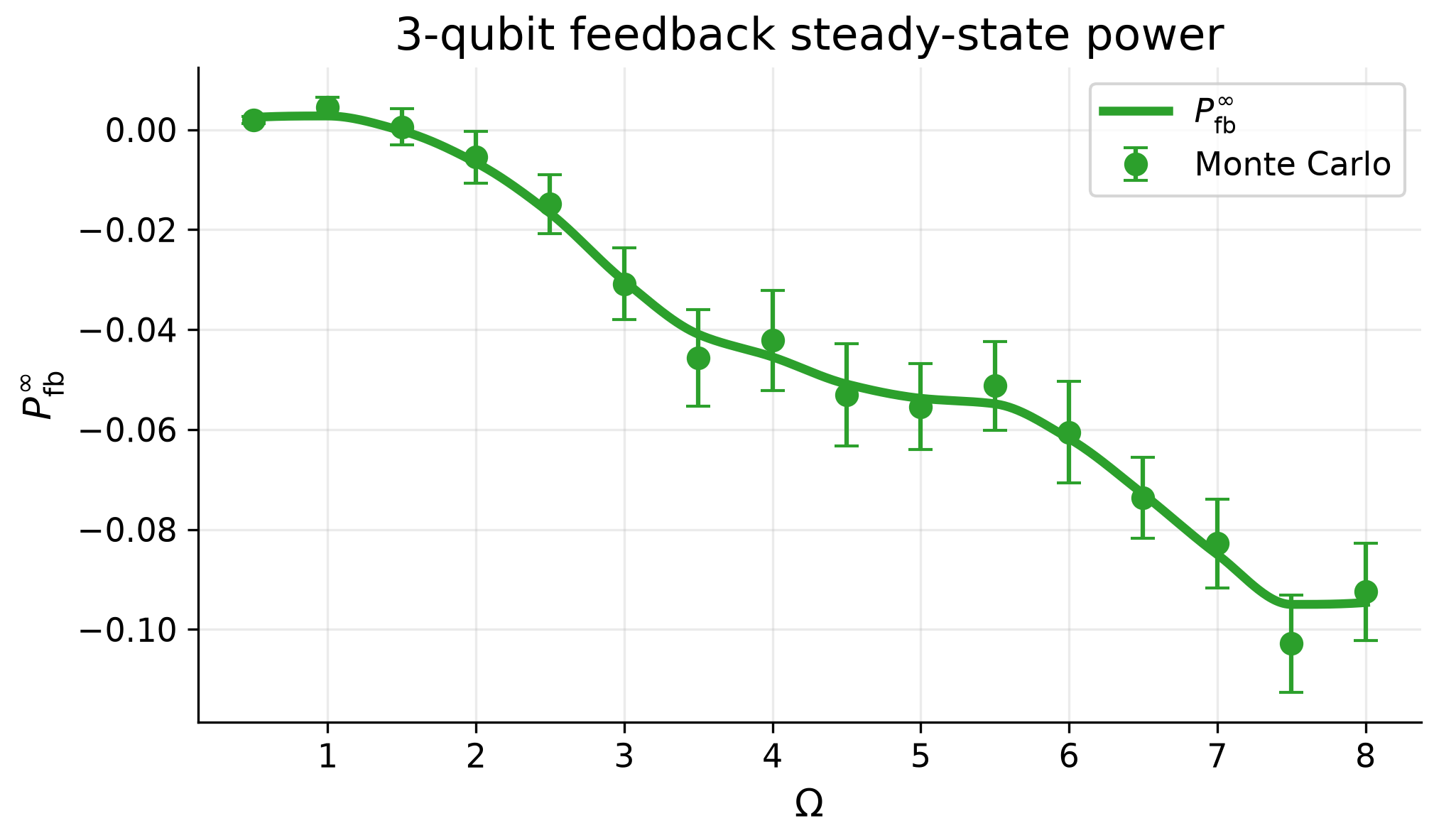}
\end{minipage}\hfill
\begin{minipage}[t]{0.46\textwidth}
    \centering
\includegraphics[width=\linewidth]{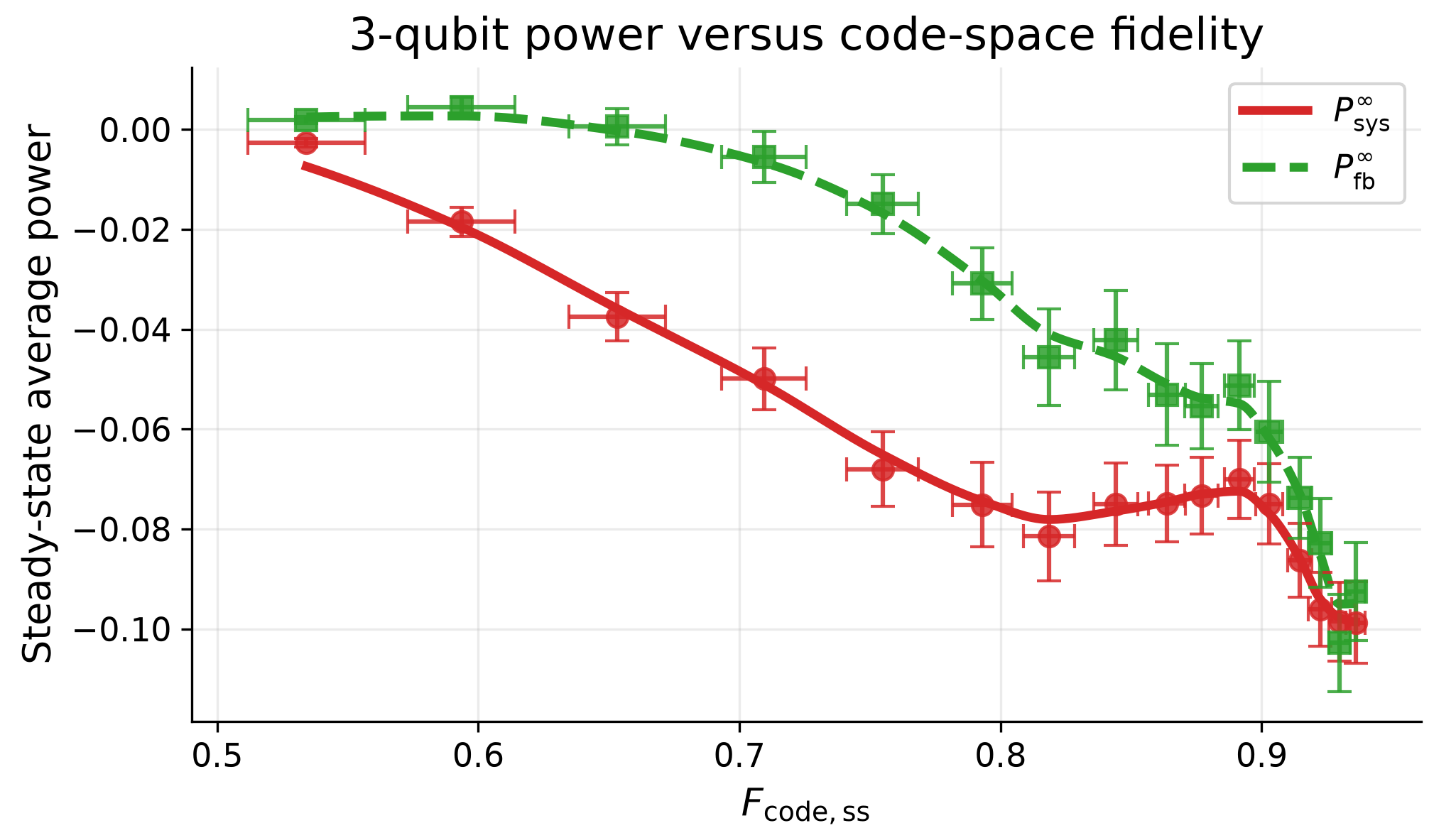}
\end{minipage}

\vspace{0.4cm}

\begin{minipage}[t]{0.46\textwidth}
    \centering
\includegraphics[width=\linewidth]{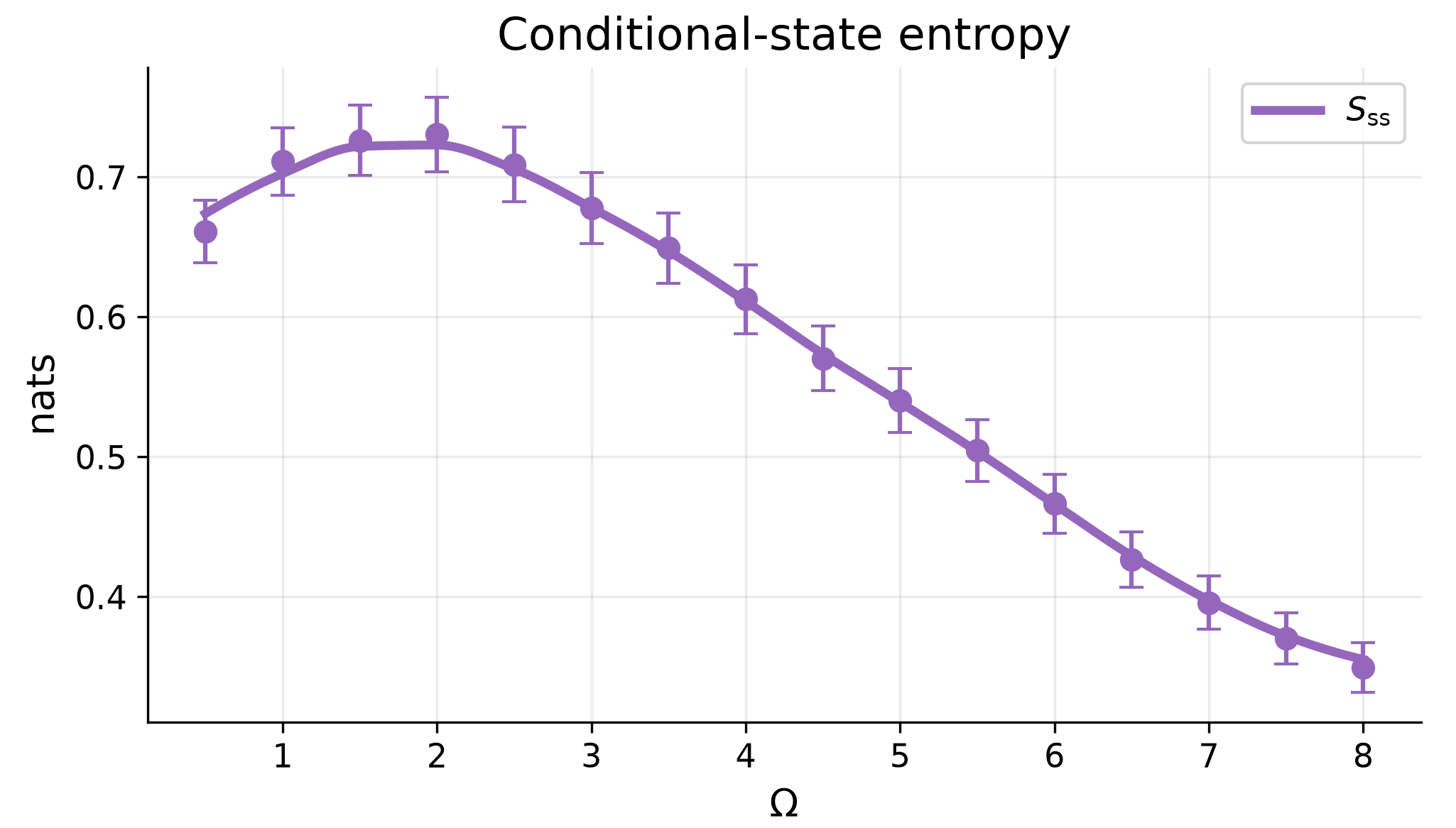}
\end{minipage}

\caption{Thermodynamic properties of the three-qubit repetition code with ADL continuous feedback protocol, as a function of the feedback strength.
(a) Logical codespace fidelity, (b) asymptotic average system power $P_{\rm sys}^{\infty}(T)$, (c) feedback power $P_{\rm fb}^{\infty}(T)$, (d) power cost versus logical fidelity and (e) von Neumann entropy.
}
\end{figure*}

\begin{subsection}{ADL feedback law}
The three-qubit ADL controller uses the syndrome-dependent gradient correlators
\begin{subequations}
\begin{align}
G_1&=Y_1Z_2+Y_1Z_3\\
G_2&=Y_2Z_1+Y_2Z_3\\
G_3&=Y_3Z_1+Y_3Z_2.
\end{align}
\end{subequations}
Their conditional expectation values are
\begin{equation}
    g_a(t)=\langle G_a \rangle
\end{equation}
for $a=1,2,3.$ The controller then applies bang-bang signs
\begin{equation}
    s_a(t)=\mathrm{sgn}(g_a(t)), 
\end{equation}
with a small hysteresis window in the numerical implementation to suppress chatter near zero. The feedback Hamiltonian is
\begin{equation}
    H_{\rm fb}(t)=\frac{\Omega}{2}(s_1X_1+s_2X_2+s_3X_3).
\end{equation}

Because the stabilizer measurements and feedback corrections do not
distinguish the logical basis states, the same controller applies to
any encoded state $\alpha|0_L\rangle+\beta|1_L\rangle$, up to the
logical phase evolution generated by $H_S$.

Compared with the one-qubit protocol, the control is now collective and distributed: each physical qubit is rotated by its own bounded local field, and the choice of sign for each channel is determined by the syndrome-dependent gradient observables. This is the direct three-qubit analogue of the ADL feedback law.
\end{subsection}

\begin{subsection}{Thermodynamic observables}

The thermodynamic quantities introduced in the one-qubit section carry over directly.

The internal energy \cref{eq:Int_Ene} becomes
\begin{equation}
    U(t)=\mathrm{Tr}[H_S \rho(t)]=\frac{\Delta}{2}\sum_{i=1}^3(1-\langle Z_i \rangle).
\end{equation}

The system-side power is defined by \cref{eq:P_sys} and the feedback work is given by \cref{eq:P_fb}. 

As in the one-qubit case, $W_{\rm sys}$ quantifies the energy exchanged between the feedback field and the physical qubits through noncommutativity, whereas $W_{\rm fb}$ quantifies the energetic cost of explicitly changing the feedback Hamiltonian itself. In the three-qubit code, however, both quantities now describe collective energetic resources required to stabilize an encoded logical qubit rather than a single physical one.

The von Neumann entropy is unchanged in form \cref{eq:entropy}, but now with the $8\times 8$ density matrix. 

One important difference with the one-qubit code is that in the repetition code the feedback is collective and syndrome-dependent. Instead of a single bounded control field $u(t)\sigma_x$, the controller applies three bounded local rotations weighted by the signs of the three gradient correlators $G_i$ for $i=1,2,3$. This is the mechanism by which stabilizer information is converted into active correction on the physical qubits.
\\
\end{subsection}

\begin{subsection}{Numerical Results}

The thermodynamic behavior of the three-qubit repetition code is summarized in Fig. 2. Figure 2(a) shows that the codespace population increases over the investigated range of feedback strengths.

The corresponding energetic cost is shown in Figs. 2(b) and 2(c). Both the asymptotic average system power and the feedback power increase in magnitude with the feedback strength, demonstrating that stronger logical protection requires a larger energetic investment. The fluctuations are larger than in the one-qubit case due to the stochastic syndrome dynamics and the collective feedback acting on three physical qubits, with three different gradient correlators.

Figure 2(d) directly relates the energetic cost to the achieved code-space fidelity. Similar to the one-qubit model, the first improvements in fidelity are obtained with relatively little power, whereas approaching perfect logical protection requires increasingly larger energetic resources.

Finally, Fig.~2(e) shows the steady-state conditional von Neumann
entropy. Its reduction at larger feedback strengths indicates that
continuous syndrome monitoring and corrective feedback reduce the
uncertainty remaining in the conditional state. This quantity should
not be interpreted as the entropy stored in the classical controller
memory.
    
\end{subsection}
\end{section}

\begin{section}{Conclusion}\label{sec:conclusion}

We have developed a thermodynamic description of measurement-based continuous quantum error correction. The framework distinguishes the system-side power associated with energy transfer between the feedback field and the system Hamiltonian $P_{\rm sys}$ from the controller-side power required to modulate the feedback Hamiltonian $P_{\rm fb}$. These quantities capture different energetic aspects of continuous stabilization and complement conventional performance measures based exclusively on fidelity.

We applied the framework first to the one-qubit ADL protocol and then to the three-qubit repetition code under continuous stabilizer monitoring. 

In both cases, increasing the feedback strength improves steady-state stabilization. The conditional-state entropy decreases monotonically in the one-qubit model, whereas the three-qubit model exhibits an initial increase followed by a reduction at larger feedback strengths.

However, the fidelity eventually enters a saturation regime while the magnitudes of the system-side and feedback powers continue to increase. Approaching perfect stabilization, fidelity closer to one, requires progressively larger energetic resources for increasingly smaller improvements in performance.

These results reveal a relationship between logical stabilization,
feedback-induced energy transfer, and the uncertainty remaining in the
conditional quantum state and establish average feedback power as a relevant physical resource for evaluating continuous controllers. The proposed description provides a basis for comparing error-correction strategies according to both their protective performance and their energetic requirements. Future work may incorporate these quantities directly into power-constrained feedback laws and extend the analysis to larger stabilizer codes, bosonic codes, and experimentally realistic finite-bandwidth controllers. 

Future work will also investigate optimized feedback Hamiltonians that
minimize the required power while maximizing the rate of fidelity
improvement.

\end{section}

\begin{acknowledgments}
We thank Luis Pedro Garc\'ia-Pintos, Brian Barch, Todd Brun, Aziz Fall, Anirudh Lanka, Michele Minervini, Leesok Kim, Aar\'{o}n Villanueva and Hui Khoon Ng for helpful conversations about this work and related topics. LJ is supported by a Feynman Fellowship through LANL LDRD project number 20230788PRD1. PBM acknowledges the support of the Brazilian agency CNPq for funding support, grant No. 140264/2026-4. This work was supported by the U.S. Department of Energy (DOE) through a quantum computing program sponsored by the Los Alamos National Laboratory (LANL) Information Science \& Technology Institute. JGN and PBM acknowledge the Los Alamos Quantum Computing Summer School (QCSS), during which part of this work was developed.
\end{acknowledgments}

\bibliography{references} 

\end{document}